\documentclass[prd,aps,twocolumn,amsmath,amssymb,nofootinbib,preprintnumbers]{revtex4-1}

\usepackage{graphicx}
\usepackage{dcolumn}
\usepackage{bm}
\usepackage{amsmath}
\usepackage{amsfonts}
\usepackage{ifthen}

\def\ls{\mathrel{\lower4pt\vbox{\lineskip=0pt\baselineskip=0pt
           \hbox{$<$}\hbox{$\sim$}}}}
\def\gs{\mathrel{\lower4pt\vbox{\lineskip=0pt\baselineskip=0pt
           \hbox{$>$}\hbox{$\sim$}}}}
\def\drawbox#1#2{\hrule height#2pt

\hbox{\vrule width#2pt height#1pt \kern#1pt
              \vrule width#2pt}
              \hrule height#2pt}

\def\Asym#1#2{\vcenter{\vbox{\drawbox{#1}{#2}
              \kern-#2pt       
              \drawbox{#1}{#2}}}}

\def\beq{\begin{equation}
\def\eeq{\end{equation}}}

\newcommand{\be}{\begin{equation}}
\newcommand{\ee}{\end{equation}}
\newcommand{\bea}{\begin{eqnarray}}
\newcommand{\eea}{\end{eqnarray}}

\begin{document}

\title{Constructing Flat Inflationary Potentials in Supersymmetry}

\author{Rouzbeh Allahverdi$^{1}$}
\author{Sean Downes$^{2}$}
\author{Bhaskar Dutta$^{2}$}

\affiliation{$^{1}$~Department of Physics and Astronomy, University of New Mexico, Albuquerque, NM 87131-0001, USA \\
$^{2}$~Department of Physics and Astronomy, Texas A\&M University, College Station, TX 77843-4242, USA}

\begin{abstract}
We show that in supersymmetry one can obtain inflationary potentials in the observable sector that are sufficiently flat at sub-Planckian field values. Structure of the supersymmetric scalar potential along a flat direction combined with the existence of higher order terms in an effective field theory expansion allows one to find scales below the effective field theory cut off where two or a higher number of the potential derivatives may vanish. As an explicit example, we demonstrate that inflection point inflation within a broad range of scales ${\cal O}({\rm TeV}) \ll H_{\rm inf} < 3 \times 10^9$ GeV can be accommodated within weak scale supersymmetry. The fine tuning of model parameters needed for successful inflation is considerably improved in this scenario.
\end{abstract}
MIFPA-11-23 \\ June, 2011
\maketitle



Inflation is the dominant paradigm in early universe cosmology to solve the problems of the hot big-bang model and create the seeds for structure formation. Although observations strongly support a period of inflation~\cite{WMAP}, a natural embedding of inflation within particle physics~\cite{MR} remains a challenge. At the heart of the problem is identifying the inflaton with a scalar field that has a natural place in particle physics and has a sufficiently flat potential that is not destroyed by unknown effects above the Planck scale.

Attempts have been made in recent years to realize inflation in realistic extensions of the Standard Model. In particular, it has been shown that inflation can be successfully embedded within the Minimal Supersymmetric Standard Model (MSSM)~\cite{AEGM,AEGJM} and its simple extensions~\cite{AKM}.
If appropriate relations hold between the supersymmetry (SUSY) breaking parameters, one finds a point of inflection in the scalar potential along $D$-flat directions~\cite{GKM} in these models.
Inflation occurs in the vicinity of the inflection point. For specific flat directions, it generates acceptable density perturbations and leads to successful post-inflationary cosmology~\cite{ADM1,ADM2,ADS,AFGM}.
For weak scale SUSY, the Vacuum Expectation Value (VEV) of the inflection point is a few orders of magnitude below $M_{\rm P}$. The sub-Planckian field value makes the model rather insensitive to the details of an ultraviolet completion. On the other hand, it implies a severe fine tuning between the SUSY breaking parameters in order to have successful inflation (for various aspects of this problem, see~\cite{ADS,EMS}). The fine tuning gets alleviated in high scale SUSY where the inflection point has a larger VEV, but this is not appealing from the point of view of particle physics phenomenology.

In this letter we provide a new and general prescription for constructing potentials in the observable sector that are suitable for inflation at sub-Planckian VEVs. 
We show that the structure of the SUSY preserving part of the scalar potential alone allows us to find points along a flat direction where two or more of the potential derivatives may vanish. Successful inflation can occur around these points within a broad range of scales. These points can be very close to the Planck scale even for TeV scale SUSY, a byproduct of which is considerable amelioration of the fine tuning problem.

We start by considering the superpotential $W$ and scalar potential $V$ for a $D$-flat direction $\phi$ that is represented by a cubic gauge-invariant monomial:
\begin{eqnarray}
W(\phi) & = & \sum_{n}{{\lambda_n \over 3n} {\phi^{3n} \over M^{3n-3}_{\rm P}}} \, , \label{supot} \\
V(\phi) = \vert f(\phi) \vert^2 ~ & , & ~ f(\phi) \equiv {d W(\phi) \over d \phi} \, . \label{scpot}
\end{eqnarray}
The lowest order term ($n=1$) is a typical Yukawa coupling (in MSSM or beyond),
and higher order terms ($n > 1$) are induced by new physics at high scales. If we take this scale to be $M_{\rm P}$, then $\lambda_n \leq 1$ and the effective field theory expansion is valid at $\vert \phi \vert \ll M_{\rm P}$ (taken at least an order of magnitude below $M_{\rm P}$).

%
%
First, we consider the case where the first three terms are dominant:
\begin{eqnarray} \label{n=3}
f(\phi) = \lambda_1 \phi^2 + \lambda_2 {\phi^{5} \over M^3_{\rm P}} + \lambda_3 {\phi^{8} \over M^{6}_{\rm P}} \, .
\end{eqnarray}
%
%
%
Then $V^{\prime}(\phi) = f^{\prime}(\phi) f^*(\phi) + {\rm h.c.}$ vanishes at the points $\phi = (0,~a^{1/3},~b^{1/3}) M_{\rm P}$, where $a + b = - 5 \lambda_2/8 \lambda_3$ and $ab = \lambda_1/4 \lambda_3$. These solutions exist for any values of $\lambda_{1,2,3}$ since $\phi$ is complex.
%
%

If $a=b$, we will have $\partial V/\partial \phi = \partial V/\partial \phi^* = \partial^2 V /{\partial \phi}^2 = \partial^2 V/{\partial \phi^*}^2 = 0$. This happens when
%
\beq \label{infcond}
\lambda^2_2 = {64 \over 25} \lambda_1 \lambda_3.
\eeq
We note that phase of $\lambda_1$ can be always rotated away, while the relative phase between $\lambda_1,~\lambda_2$ can be absorbed by a redefinition of $\phi$. Therefore we can choose $\lambda_1$ and $\lambda_2$ to be real and positive, and then the above condition implies that $\lambda_3 > 0$.
For $\lambda_{1,2,3} > 0$, the first and second derivatives of the potential vanish along both the radial and angular direction in the complex $\phi$ plane at:
\beq \label{infloc}
\phi = \phi_0 ~ {\rm exp}[i {\pi \over 3},~i \pi, i {5 \pi \over 3}] ~ ~ , ~ ~ \phi_0 = \left({5 \over 16} {\lambda_2 \over \lambda_3}\right)^{1/3} M_{\rm P} .
\eeq
%
The validity of this result
within the effective field theory expansion is ensured if $\phi_0 \ll M_{\rm P}$, which leads to the condition
%
$\lambda_2 \ll \lambda_3 \ls 1$.
%

Focusing on the radial direction, the dynamics of inflation is governed by the potential $V_0 = 81 \lambda^2_1 \phi^4_0/400$ and its 3rd derivative $V^{\prime \prime \prime}_0 = 162 \lambda^2_1 \phi_0/5$ at these points. The slow-roll conditions for inflation are satisfied within an interval $\Delta \phi \sim \phi^3_0/80 M^2_{\rm P}$ around $\phi_0$.

The amplitude of observationally relevant density perturbations generated during inflation is given by $\delta_H = V^{\prime \prime \prime}_0 {\cal N}^2_{\rm COBE}/30 \pi H_{\rm inf}$~\cite{AEGJM}. Here $H_{\rm inf} =(V_0/3 M^2_{\rm P})^{1/2}$ is the Hubble expansion rate during inflation, and ${\cal N}_{\rm COBE} = 66.9 +(1/4){\rm ln}(V_0/M^4_{\rm P})$~\cite{LL} (assuming rapid transition from inflation to a radiation-dominated universe, which is the case when the inflaton is a MSSM flat direction~\cite{AFGM}). Obtaining the correct amplitude for perturbations
requires that
\beq \label{pert}
\lambda_1 \left({16 \over 5} {\lambda_3 \over \lambda_2}\right)^{1/3} \sim 10^{-8} .
\eeq
This gives rise to an absolute upper bound $\lambda_1 < 10^{-8}$, which is saturated when $\phi_0 \rightarrow M_{\rm P}$, leading to an upper bound of $H_{\rm inf} < 3 \times 10^{9}$ GeV.

Satisfying all of the conditions for a successful inflation requires small values for $\lambda_{1,2,3}$. In Figure 1 we show the allowed region of the parameter space by projecting to the $\lambda_3/\lambda_1-\lambda_2/\lambda_1$ plane. The solid line satisfies the condition given in~(\ref{infcond}). The dashed lines show the condition in~(\ref{pert}) for $\lambda_1 = 1 \times 10^{-8}, ~ 5 \times 10^{-9} , ~ 1 \times 10^{-9}$ respectively (from bottom to top). From the figure, one acceptable choice of parameters is: $\lambda_1 \sim 10^{-9},~\lambda_2 \sim 10^{-6},~\lambda_3 \sim 10^{-3}$. This results in $\phi_0 \sim 10^{-1} M_{\rm P}$. $\phi_0$ becomes closer to $M_{\rm P}$ as we move down along the solid line.

One can see that $n \geq 4$ terms in~(\ref{supot},\ref{scpot}) are negligible around $\phi_0$ (with the contribution from the $n=4$ term being marginal) for $\lambda_n \sim {\cal O}(1)$.
It is also seen that the soft mass term and the $A$-terms associated with the $n \leq 3$ terms in Eq.~(\ref{supot}) make tiny contributions to the potential around $\phi_0$, and hence can be neglected, for TeV scale SUSY. As a matter fact, the contribution of any mass term $\leq 10^8$ GeV will be insignificant.

Deviations from the relation in Eq.~(\ref{infcond}) will result in a non-vanishing $V^{\prime}$ along the radial direction
. Parameterizing the deviation as
%
$\alpha \equiv 1 - (25 \lambda^2_2/128 \lambda_1 \lambda_3)$,
%
we find that $\phi_0 \rightarrow \phi_0 - (2/9) \alpha \phi_0$ and $V^{\prime}_0 = (9/5) \alpha \lambda^2_1 \phi^3_0$.
Obtaining a scalar spectral index $n_s < 1$ requires that
%
$0 < \alpha \ll {\cal N}_{\rm COBE}^{-2} (\phi_0/M_{\rm P})^4$~\cite{AEGJM,LK}. A considerable improvement in the fine tuning is evident, as compared with MSSM inflation, since much larger values of $\phi_0$ are found in this case. For the above choice of parameters one has $\phi_0 \sim 10^{-1} M_{\rm P}$, which is larger by three orders of magnitude than that in the case of the MSSM inflation~\cite{AEGJM}, and the resulting $\alpha$ is 12 orders of magnitude larger.

\begin{figure}[tbp]
%
\includegraphics[viewport = 0 10 290 220,width=8.2cm]{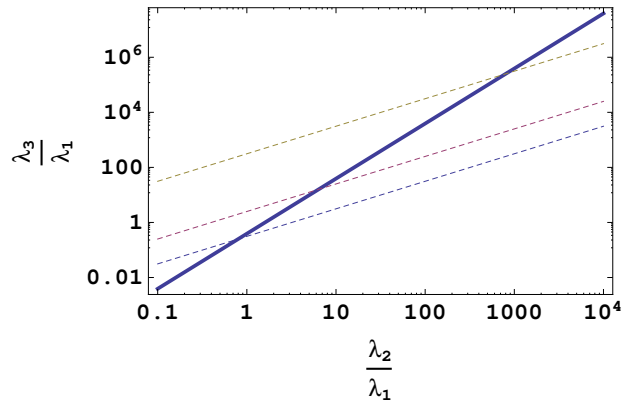}
%
\caption{
The solid line corresponds to the condition in Eq.~(\ref{infcond}). The dashed lines. The dashed represent $\lambda_1 = {\rm const.}$ contours corresponding to density perturbations condition~(\ref{pert}) for $\lambda_1 = 1 \times 10^{-8}, ~ 5 \times 10^{-9}, ~ 1 \times 10^{-9}$ respectively (from bottom to top).}
\label{Fig1}
\end{figure}

Inclusion of more terms in~(\ref{supot},\ref{scpot}) provides further possibilities to obtain a flat potential. For example, if we also include the $n=4$ term:
\begin{eqnarray} \label{n=4}
f(\phi) = \lambda_1 \phi^2 + \lambda_2 {\phi^5 \over M^3_{\rm P}} +
\lambda_3 {\phi^8 \over M^6_{\rm P}} + \lambda_4 {\phi^{11} \over M^9_{\rm
P}} \, ,
\end{eqnarray}
then $V^{\prime} = 0$ at $\phi = (0,~a^{1/3},~b^{1/3},~c^{1/3}) M_{\rm P}$, where $a + b + c = - 8 \lambda_3/11 \lambda_4$, $ab + bc + ac = 5 \lambda_2/11 \lambda_4$, and $abc = - 2 \lambda_1/11 \lambda_4$. Assuming that $\vert a \vert \leq \vert b \vert \leq \vert c \vert$, we will have $V^{\prime} = V^{\prime \prime} = 0$ if $a = b$, or if $b = c$. Note, however, that for a successful scenario only the former case may be acceptable. Otherwise, the inflaton may roll toward a minimum away from the origin and settle there, instead of the true minimum at $\phi = 0$, after inflation.

The corresponding parameter space can be quantitatively understood under the scaling $\phi \rightarrow \gamma \phi$, $\lambda_4 \rightarrow \lambda_4$, $\lambda_3 \rightarrow \gamma^3 \lambda_3$, $\lambda_2 \rightarrow \gamma^6 \lambda_2$, $\lambda_1 \rightarrow \gamma^9 \lambda_1$, which results in $f^{\prime}(\phi) \rightarrow \gamma^{10} f^{\prime}(\phi)$. Then it will be sufficient to consider the cases $\lambda_3 = 0,\pm 1$.
Since the phase of $\lambda_3$ can be rotated away, it suffices to consider the cases $\lambda_3 = 0$ and $\lambda_3 = 1$. Then
%
\beq
\lambda_3 = 0 \Longrightarrow \lambda^2_1 \lambda_4 = - {125 \lambda_2^3 \over 297} ,
\eeq
with $c = -2a$, and
\beq
\lambda_3 = + 1 \Longrightarrow {2 \lambda_1 \over11 \lambda_4} = 2 a^3 + a^2 ~ , ~ {5 \lambda_2 \over 11 \lambda_4} = -3 a^2 - 2 a,
\eeq
with $c = -2 a - 1$.
In these cases one can have successful inflation about the points $\phi = a^{1/3} M_{\rm P}$ as discussed above.

In the special case that $a = b = c$, we will have $V^{\prime} = V^{\prime \prime} = V^{\prime \prime \prime} = 0$. This happens when
\beq \label{infcond2}
\lambda^2_2 = {48 \over 25} \lambda_1 \lambda_3 ~ ~ ~ , ~ ~ ~ \lambda^2_3 = {165 \over 64} \lambda_2 \lambda_4 ,
\eeq
at the points (for $\lambda_{1,2,3,4} > 0$)
\beq \label{infloc2}
\phi = \phi_0 ~ {\rm exp}[i {\pi \over 3},~i \pi, i {5 \pi \over 3}] ~ ~ ~ , ~ ~ ~ \phi_0 = \left({8 \over 33} {\lambda_3 \over \lambda_4}\right)^{1/3} M_{\rm P} .
\eeq
Successful inflation can occur around $\phi_0$, similar to the previous case, if $\lambda_1 \ll \lambda_2 \ll \lambda_3 \ll \lambda_4 \ls 1$. The condition to obtain acceptable density perturbations results in an upper bound $\lambda_1 \ls 10^{-8}$ as before.
As an interesting consequence, we can show that fine tuning in the conditions to have successful inflation, see~(\ref{infcond2}), will be $\propto {\cal N}_{\rm COBE}^{-3/2} (\phi_0/M_{\rm P})^3$ in this case. This is a significant improvement over the previously discussed case, see~(\ref{infcond}), for which the fine tuning is $\propto {\cal N}_{\rm COBE}^{-2} (\phi_0/M_{\rm P})^4$.

Again, the higher order ($n \geq 5$) terms in~(\ref{supot}) are negligible at the scale of $\phi_0$ if the corresponding couplings are $\ls {\cal O}(1)$. The contribution of SUSY breaking terms to the potential will also be tiny.

In general, including a number $n \geq 3$ of terms in the superpotential~(\ref{supot}) not only allows us to find points at which $V^{\prime} = V^{\prime \prime} = 0$, thus realizing inflection point inflation, but we can also find points where up to the first $n-1$ derivatives of the potential vanish. This is a direct consequence of the fact that $f^{\prime}(\phi) \propto \phi \Pi_{i=1}^{n-1}{[(\phi/M_{\rm P})^3 - a_i]}$, and the roots $a_i$ always exist for a complex field $\phi$. Inflection points, or higher degenerate points, arise when two or more of the roots coincide.

Mathematics of all possible inflationary solutions for the potential in~(\ref{scpot}) can be understood in the context of ``Catastrophe Theory''. There the highest order term in a function is called the {\it catastrophe germ} and the coefficients of the lower order terms are called {\it control parameters}~\cite{arnold3}. The points associated with the lower order germ can be nested when the function is extended to include higher order germs by suitably choosing the control parameters. The whole description can be described rigorously in the context of Lie Algebras. It will also allow us to handle cases with more than one flat direction. An explicit example of applying catastrophe theory to understand inflation has been presented recently in the context of racetrack models in type IIB string theory~\cite{Downes:2011gi}.

%
%
%
%

%
%
%
%

Having successful inflation about a point $\phi_0 \ll M_{\rm P}$ when a number $n$ of the terms in~(\ref{supot}) are included
results in the condition $\lambda_1 \ll ... \ll \lambda_{n} \ls 1$. Having very small coupling(s) is a generic issue in inflationary model building. For example, in the $\lambda \phi^4$ chaotic inflation, one needs to have $\lambda \sim 10^{-13}$.

Small superpotential couplings are technically natural since radiative corrections to such couplings arise from the wavefunction renormalization, and hence are proportional to the couplings themselves. Their smallness at the tree level can be attributed to a symmetry that broken at a scale $v \ll M_{\rm P}$. Terms of order $n$ can originate from terms of order $m > n$, which results in a suppression $\propto (v/M_{\rm P})^{m-n}$ after symmetry breaking. As we pointed out, the density perturbations condition sets an absolute upper bound $\lambda_1 < 10^{-8}$ on the renormalizable coupling in~(\ref{supot}). This is too small to be identified with quark or lepton Yukawa couplings, but may be easily related to the Dirac and/or Majorana Yukawa couplings of neutrinos.

One comment is in order at this point. In our discussion we have considered superpotential terms of the form $\phi^{3n}$~(\ref{supot}). This is strictly correct for $D$-flat directions that are represented by a cubic monomial, as happens for those corresponding to Yukawa couplings. Our argument, however, is quite general and applies to any superpotential that is a polynomial function of $\phi$. Also, one may have superpotential terms of the form $\phi^n \chi/M^{n+1}_{\rm P}$, if allowed by symmetries, where $\chi$ is another scalar field. These terms also make nonzero contributions to $V(\phi)$. It is possible to show that one can still find points at which $V^{\prime}, ~ V^{\prime \prime}, ~ ...$ vanish in the presence of these terms if appropriate relations hold among the corresponding couplings.

Finally, we make some comments regarding the post-inflationary universe in this scenario. After inflation, the inflaton starts oscillating about the minimum of its potential at $\phi = 0$. The situation is qualitatively similar to that in MSSM inflation where inflaton oscillations lead to particle creation via a combination of nonperturtbative and perturbative effects~\cite{AFGM}. There are quantitative differences between the two cases. First, the potential around the minimum is quartic in this case $V(\phi) \sim \lambda^2_1 \phi^4$, while in the case of the MSSM inflation it is quadratic. This can consequently lead to differences in nonperturbative particle creation~\cite{quartic} as well as their subsequent perturbative decay. Second, in the case of the MSSM inflation the inflaton energy is transferred to relativistic particles very efficiently, which in turn thermalize within a Hubble time. The efficiency of reheating in the scenario discussed here depends on the nature of the flat direction $\phi$ as well as the ratio of the frequency of inflaton oscillations, which is $\sim \lambda_1 \phi_0$, and $H_{\rm inf}$.

Another point to note is that $H_{\rm inf} \gg {\cal O}({\rm TeV})$ is typical in this scenario. As a result, the MSSM flat directions may acquire large VEVs during inflation~\cite{DRT}. This can lead to various cosmological consequences~\cite{flat}, for example, Affleck-Dine lepto/baryogenesis~\cite{AD}.


In summary, we have provided a prescription for a systematic construction of flat inflationary potentials at sub-Planckian field values in the observable sector within supersymmetry. Structure of the supersymmetric scalar potential allows us to obtain points along a flat direction where any number of potential derivatives may vanish. Successful inflation can occur around these points within a broad range of scales ${\cal O}({\rm TeV}) \ll H_{\rm inf} < 3 \times 10^9$ GeV, with the scale of SUSY breaking kept around TeV. As a consequence, this leads to a considerable amelioration of the fine tuning of flat direction inflation. The validity of the construction presented here within the effective field theory expansion requires that some of the superpotential couplings be  small. This smallness is technically natural and its origin may be related to the neutrino sector.\\
\\

\noindent
{\it Acknowledgements:} This work is supported in part by the DOE grant DE-FG02-95ER40917 and by the University of New Mexico Office of Research. R.A and B.D wish to thank the hospitality by CERN Theoretical Physics Division where this work was completed.

\end{document}